\documentclass[twocolumn]{aastex63}
\usepackage[T1]{fontenc} 
\usepackage{newpxtext}
\usepackage{newpxmath}

\usepackage{soul}

\newcommand{\uat}[2]{\href{http://astrothesaurus.org/uat/#2}{#1 (#2)}}

\providecommand{\sorthelp}[1]{}
\usepackage{xspace}
\newcommand{\ud}{\,\mathrm d}
\DeclareMathOperator{\Tr}{Tr}
\newcommand{\wmap}{\textsl{WMAP}\xspace}
\newcommand{\planck}{\textsl{Planck}\xspace}
\newcommand{\EE}{$\hat C_\ell^\mathrm{EE}$\xspace}
\newcommand{\TE}{$\hat C_\ell^\mathrm{TE}$\xspace}
\newcommand{\TT}{$\hat C_\ell^\mathrm{TT}$\xspace}
\newcommand{\TTTEEE}{${\hat C_\ell^\mathrm{TT}+\hat C_\ell^\mathrm{TE}+\hat C_\ell^\mathrm{EE}}$\xspace}

\newcommand{\covC}{\boldsymbol{\mathsf C}}
\newcommand{\covChat}{\hat{\boldsymbol{\mathsf C}}}

\received{October 1, 2019}
\revised{December 3, 2019}
\accepted{\today}
\submitjournal{ApJ}

\shorttitle{Reionization Histories}
\shortauthors{Watts et al.}

\usepackage[utf8x]{inputenc}
\usepackage{graphicx}
\usepackage{amsmath}
\usepackage{enumitem}

\begin{document}

\title{Beyond optical depth: Future determination of ionization history from the CMB}

\author[0000-0002-5437-6121]{D.~J.~Watts}
\affil{JHU Department of Physics and Astronomy, 3701 San Martin Drive, Baltimore MD, 21218, USA}

\author[0000-0002-2147-2248]{G.~E.~Addison}
\affil{JHU Department of Physics and Astronomy, 3701 San Martin Drive, Baltimore MD, 21218, USA}

\author[0000-0001-8839-7206]{C.~L.~Bennett}
\affil{JHU Department of Physics and Astronomy, 3701 San Martin Drive, Baltimore MD, 21218, USA}

\author[0000-0003-3017-3474]{J.~L.~Weiland}
\affil{JHU Department of Physics and Astronomy, 3701 San Martin Drive, Baltimore MD, 21218, USA}

\correspondingauthor{Duncan J. Watts}
\email{dwatts@jhu.edu}

%\keywords{cosmic background radiation --- cosmological parameters --- reionization}


\begin{abstract}
We explore the fundamental limits to which reionization histories can be constrained using only large-scale cosmic microwave background (CMB) anisotropy measurements.
The redshift distribution of the fractional ionization $x_e(z)$ affects the angular distribution of CMB polarization. 
We project constraints on the reionization history of the universe using low-noise full-sky temperature and E-mode measurements of the CMB.
We show that the measured TE power spectrum, \TE, has roughly one quarter of the constraining power of \EE on the reionization optical depth $\tau$, and its addition improves the precision on $\tau$ by 20\% over using \EE only. 
We also use a two-step reionization model with an additional high-redshift step, parameterized by an early ionization fraction $x_e^\mathrm{min}$,  and a late reionization step at $z_\mathrm{re}$.
We find that future high signal-to-noise measurements of the multipoles $10\leqslant\ell<20$ are especially important for breaking the degeneracy between $x_e^\mathrm{min}$ and $z_\mathrm{re}$. In addition, we show that the uncertainties on these parameters determined from a map with sensitivity $10\,\mathrm{\mu K\,arcmin}$ are less than 5\% larger than the uncertainties in the noiseless case, making this noise level a natural target for future large sky area E-mode measurements. 
\end{abstract}

\section{Introduction}
\label{sec:intro}

\begin{figure}
    \centering
    \includegraphics[width=\columnwidth]{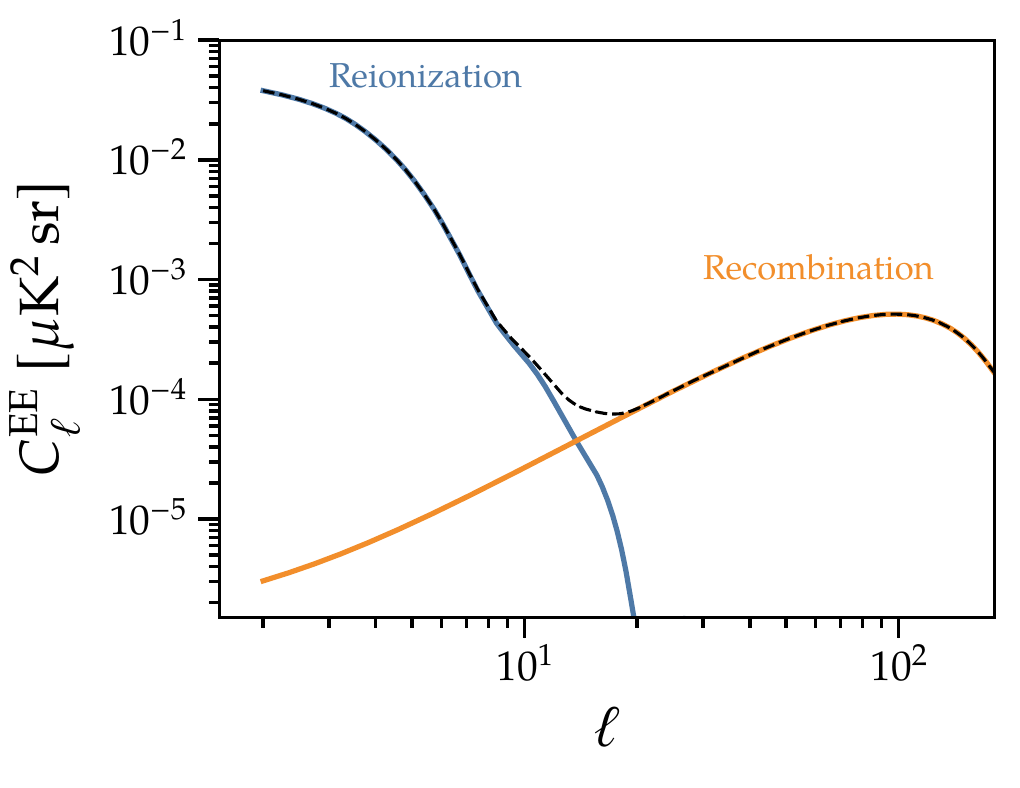}
    \caption{Effect of reionization on the $C_\ell^\mathrm{EE}$ power spectrum. 
    We take the difference between an E-mode signal with $\tau=0$ and one with $\tau=0.06$ with fixed $A_s e^{-2\tau}$ to demonstrate the effects of tanh-like reionization on $C_\ell^\mathrm{EE}$ versus those from recombination. The black dashed line is the total $C_\ell^\mathrm{EE}$ spectrum when $\tau=0.06$.
    The E-mode signal from recombination dominates above $\ell\gtrsim20$, whereas the reionization signal emerges at multipoles $\ell\lesssim20$.}
    \label{fig:reio_reco}
\end{figure}

Cosmic reionization is a poorly understood part of standard $\Lambda$CDM cosmology.
Reionization, when  neutral hydrogen and helium in the intergalactic medium (IGM) become ionized,  creates a plasma that scatters cosmic microwave background (CMB) photons~\citep{rees, basko, bond}. This reduces the amplitude of the CMB anisotropy at angular scales $\ell\gtrsim10$ and creates additional polarized power that dominates at scales $\ell\lesssim10$~\citep{zaldarriaga}.
We illustrate the separate effects of reionization and recombination on the E-mode power spectrum in \autoref{fig:reio_reco}.
Because the temperature and E-mode polarization angular power spectra ($C_\ell^\mathrm{TT}$, $C_\ell^\mathrm{TE}$, and $C_\ell^\mathrm{EE}$) depend on the redshift of scattering, their characterization at high signal-to-noise can be used to constrain ionization histories.

It is observationally known that after the universe became neutral at the epoch of recombination, by $z=6$ it was ionized once again~\citep[e.g.,][]{gunnpeterson,   becker, Fan2006}. 
Determinations of the ionization fraction of the IGM have been made at redshifts $6\lesssim z\lesssim 8$~\replaced{\citep{bouwens, greig, banados, Mason2018, davies}}
{\citep{bouwens, McGreer2015,Greig2017,banados, davies,Mason2018, Greig2019,Mason2019}}
by probing the epoch of reionization via measurements of Lyman $\alpha$ emission, but these data are sparse, and do not yet constrain the free electron fraction during the epoch of reionization~\citep[see, e.g.,][Figure 36]{plancklegacy}.

Commonly, CMB constraints on the reionization history of the universe are derived assuming a sharp transition from a neutral to fully ionized IGM.
Measurements of the large-scale CMB polarization constrain the ionization history by inferring the optical depth to the last scattering surface of the CMB, $\tau\equiv \int_{t_\mathrm{lss}}^{t_0} c\sigma_\mathrm Tn_e(t)\ud t$, where $c$ is the speed of light, $\sigma_\mathrm T$ is the Thomson scattering cross section, $n_e(t)$ is the free electron number density, $t_0$ is the current age of the universe, and $t_\mathrm{lss}$ is the last time photons interacted with matter during the epoch of recombination. Determining the free electron density $n_e(t)$ is then an inverse problem that relies on assumptions and priors. For example, a  tanh-like reionization history~\citep[e.g.,][ Equation (B3)]{tanh} with a transition from neutral to ionized at a single reionization redshift $z_\mathrm{re}$ with width $\delta z_\mathrm{re}=0.5$ has been used~\citep[e.g.,][Section 3.3]{wmapparams, planckparams18}.
Observations from the \textsl{Wilkinson Microwave Anisotropy Probe} (\wmap) satellite were used to make a measurement of the optical depth from the surface of last scattering 
$\tau=0.089\pm0.014$~\citep{wmapparams}, although this decreases to $\tau=0.067\pm 0.013$ when using \planck~353 GHz data as a template to remove Galactic dust emission~\citep{plancklike16}. \citet{plancklegacy} increased the precision of this measurement to $\tau=0.0544\pm0.0073$.
\citet{pagano} claim to have further reduced large-scale \planck systematics, reporting 
$\tau=0.059\pm0.006$.

As a cross-check, it is possible to obtain competitive constraints without using CMB polarization. \planck\ temperature measurements combined with \planck\ weak lensing and baryon acoustic oscillation (BAO) data give ${\tau=0.067 \pm 0.016}$~\citep{planck2014-a15}, consistent with results using \wmap\ temperature, \planck\ weak lensing, and BAO data, ${\tau=0.066\pm0.020}$~\citep{weilandtau}.
\citet{weilandtau} include a compilation of $\tau$ measurements, and conclude that the measured values are all consistent with $\tau=0.07\pm0.02$.
Unlike the Hubble constant $H_0$,~
(e.g., \citealt{bernal}, \citealt{freedman}, \citealt{addisonH0}, and \citealt{riess}),
the issue with reionization is not tension between measurements, 
but a lack of desired precision.

Using the one-to-one mapping of $\tau\leftrightarrow z_\mathrm{re}$ in tanh-like reionization, \citet{plancklegacy} use the low-$\ell$ polarization power spectra to infer ${z_\mathrm{re}=7.67\pm0.73}$ (\planck likelihood \texttt{Plik} best fit), while measurements of the kinetic Sunyaev--Zel'dovich effect at arcminute scales by the South Pole Telescope (SPT) and the Atacama Cosmology Telescope (ACT) can be used to limit the duration of inhomogeneous reionization to ${\delta z_\mathrm{re}<2.8}$ at the 95\% C.L. with the prior that reionization ends by $z=6$~\citep{sptksz, actksz, planckreio}.

\added{While the CMB probes the reionization process through the scattering of photons by free electrons, direct observations of the neutral hydrogen fraction would constrain the ionization state of the universe during the epoch of reionization directly. The EDGES high-band experiment \citep{Monsalve2017} measures  the brightness temperature of the global 21~cm emission from neutral hydrogen, $T_{21}$, at frequencies 90--190~MHz (redshifts 6.5--14.8).
\citet{Monsalve2017} assume a hot IGM with spin temperature $T_\mathrm s=28~\mathrm K\gg T_\mathrm{CMB}$ and find a lower limit of $\delta z_\mathrm{re}\gtrsim 0.5$ with 95\% confidence over the redshift range $6.6\lesssim z\lesssim 11$. The brightness temperature has an rms scatter of 17~mK, which corresponds to a $x_\textsc{Hi}$ rms scatter of $\gtrsim0.5$, assuming a fixed spin temperature $T_\mathrm s=28~\mathrm{mK}$. With a future rms sensitivity $\sim1~\mathrm{mK}$ measurement of $T_{21}$, a percent-level determination of $x_\textsc{Hi}(z)$ will be possible.}

\deleted{It is typically assumed that the universe was ionized by ultraviolet photons from massive stars escaping from galaxies. However, indirect measurements
using absorption spectra from gamma-ray bursts have
been made that suggest either that star formation and
gamma-ray bursts are somehow decoupled (Fruchter
et al. 2006), that the escape fraction of star forming galaxies is $\lesssim1$\% rather than the 10--20\% required to ionize
the IGM
or that the nature of star
forming galaxies changes significantly at $z \gtrsim 6$.}

\added{It is typically assumed that the universe was ionized by ultraviolet photons from massive stars escaping from galaxies. To confirm this, it is necessary to characterize the emissivity of galaxies at the onset of star formation, to see whether enough ionizing photons are generated and can escape from galaxies on a short enough timescale \citep{2014ARA&A..52..415M, McCandliss2019}. This requirement is fulfilled if there is a steep luminosity function with galaxies contributing down to an absolute magnitude $M_\mathrm{UV}\sim-13$ \citep{bouwens, Finkelstein2019} and that a volume-averaged escape fraction of 5--20\% is achieved \citep{2014ARA&A..52..415M,2016MNRAS.457.4051K,Finkelstein2019}. Determinations of the UV luminosity function through the onset of reionization is limited by sensitivity to galaxies up to redshift $z\lesssim 10$, and will be enabled by deep observations by the \textit{James Webb Space Telescope} \citep{robertson}.}

Other potential mechanisms with different redshift dependence have also been put forward. In particular, binary black hole collisions can be a source of X-rays at $z\gtrsim30$, which can raise the ionizing fraction with less fractional contribution from star formation~\citep{2016MNRAS.461.2722I}. Quasars and annihilating particles have also been proposed as ionizing mechanisms~\citep{2008mgm..conf..979M, 2015ApJ...813L...8M, 2016MNRAS.457.4051K, 2018MNRAS.473.1416M}.

As we look to the future with more sensitive data, we would like to make quantitative statements about a more detailed physical model for reionization. We explore the potential to make these constraints in this paper.
In this work, we explore potential future CMB constraints on the reionization history as parameterized by both instantaneous and extended redshift scenarios.
We focus specifically on a reionization history that consists of the usual instantaneous reionization and a second early high-redshift period of reionization that partially ionizes the universe.

This paper is organized as follows.
In \autoref{sec:emodes}, we quantify the relative constraining power for parameter likelihoods
based on \EE\ alone, \TE\ alone, and  \TTTEEE.  We define the different
likelihoods in \autoref{subsec:like_ps} and obtain constraints on the nearly instantaneous tanh-like reionization model in \autoref{subsec:like_zreio}. 
In \autoref{sec:fisher} we explore a toy reionization history model
that consists of the usual instantaneous reionization and a second early (high-redshift) period of reionization that partially ionizes the universe.
We then quantify the projected limits the CMB can impose on a reionization history of this type with free parameters of reionization redshift $z_\mathrm{re}$ and high-redshift ionization fraction $x^\mathrm{min}_e$.
We describe this modification to the standard reionization history in \autoref{subsec:reio}. We then forecast sensitivity to this model’s parameters as a function of noise and multipole range in \autoref{subsec:fisher_forecasts}, and demonstrate that most of the parameter space can be precisely constrained with the map sensitivity  ${w_p^{-1/2}\lesssim 10\,\mathrm{\mu K\,arcmin}}$ using the multipole range ${10\lesssim\ell\lesssim20}$.
We summarize our findings in \autoref{sec:conclusions}.

Throughout this paper our default model is flat $\Lambda$CDM with the \citet{planckparams18} \texttt{Plik} TT,TE,EE+lowE+lensing mean parameters ${\Omega_bh^2=0.02237}$, ${\Omega_ch^2=0.1200}$, ${100\theta_\mathrm{MC}=1.04092}$, $\ln(10^{10}A_s e^{-2\tau})=2.9352$,  and ${n_s=0.9649}$. When $\tau$ is varied, $\ln(10^{10}A_s)$ is set to $2.9352+2\tau$.

\newpage
\section{Maximizing information Used\\in power spectrum analysis}
\label{sec:emodes}

In this section, we develop a formalism for extracting reionization information from a full-sky map of the intensity and linear polarization of the CMB. In \autoref{subsec:like_ps}, we define the three likelihoods we use for different subsets of data; Wishart (for \TTTEEE), $\chi^2$ (for \EE), and variance-gamma (for \TE). In \autoref{subsec:like_zreio}, we characterize these likelihoods for the case of instantaneous tanh-like reionization.

\subsection{Likelihoods for power spectra}
\label{subsec:like_ps}

In standard $\Lambda$CDM, the CMB Stokes parameters ${\boldsymbol m=(I,Q,U)}$ are a realization of a Gaussian random process. The spherical harmonic transforms of these maps $\boldsymbol a_{\ell m}=(a_{\ell m}^\mathrm T,a_{\ell m}^\mathrm E,a_{\ell m}^\mathrm B)$ are therefore also Gaussian distributed. Neglecting B-modes, the  $\boldsymbol a_{\ell m}$ are distributed as a complex Gaussian
$\boldsymbol a_{\ell m}\sim\mathcal N\left(\boldsymbol 0,\boldsymbol{\mathsf  C}_\ell\right)$ with mean $\boldsymbol 0$ and covariance
\begin{equation}
\boldsymbol {\mathsf C}_\ell=\begin{pmatrix}
C_\ell^\mathrm{TT}&C_\ell^\mathrm{TE}
\\
C_\ell^\mathrm{TE}&C_\ell^\mathrm{EE}
\end{pmatrix}.
\end{equation}

As demonstrated in \citet{hamimeche}, the sample covariance matrix of measured power spectra $\hat{\boldsymbol{\mathsf C}}_\ell$ drawn from a theory covariance matrix $\boldsymbol{\mathsf  C}_\ell$ is given by a Wishart distribution,
\begin{equation}
(2\ell+1)\hat{\boldsymbol{\mathsf C}}_\ell\equiv\sum_m\boldsymbol a_{\ell m}^\dagger\boldsymbol a_{\ell m}^{\phantom{\dagger}}
\sim W_n(2\ell + 1, \boldsymbol{\mathsf C}_\ell)
\end{equation}
where $n$ is the number of dimensions in $\boldsymbol a_{\ell m}$. 
A Wishart distribution is a multivariate gamma distribution. A gamma distribution is a two-parameter probability distribution of which the $\chi^2$ distribution is a special case.
When considered as a likelihood $\mathcal L(\covC_\ell)\equiv P(\covChat_\ell|\covC_\ell)$, this is often normalized such that ${\chi^2_{\mathrm{eff},\ell}\equiv -2\ln \mathcal L(\covC_\ell)=0}$ when $\covC_\ell=\covChat_\ell$,
i.e.,
\begin{equation}
    -2\ln \mathcal L(\covC_\ell)=(2\ell+1)\Big[\Tr[\covChat_\ell \covC_\ell^{-1}]-\ln|\covChat_\ell\covC^{-1}|
    -n\Big].
    \label{eq:chi2_eff}
\end{equation}
In the single-dimensional case, this reduces to the more familiar $\chi^2$ distribution,
\begin{equation}
-2\ln\mathcal L(C_\ell)=(2\ell+1)\left[\frac{\hat C_\ell}{C_\ell}-\ln\frac{\hat C_\ell}{C_\ell} -1\right],
\end{equation}
in agreement with Equation 8 of \citet{hamimeche} when normalized such that $\ln\mathcal L=0$ when ${C_\ell=\hat C_\ell}$.

We also use the distribution of $\hat C_\ell^\mathrm{TE}$, i.e., the mean of the product of correlated Gaussian random variables. This was derived in \citet*{mangilli} and independently in \citet{corrgauss} and \citet{variancegamma}, and is given by a variance-gamma distribution (also called a generalized Laplace distribution or a Bessel function distribution) with functional form
\begin{equation}
\label{eq:vargam}
    P(\hat C_\ell^\mathrm{TE}|\boldsymbol\theta)
    =\frac{N^{(N+1)/2}|\hat{c}|^{(N-1)/2}e^{N\rho\hat c/\xi}K_{\ell}\left(\frac{N|\hat c|}\xi\right)}
    {2^{(N-1)/2}\sqrt\pi\Gamma(N/2)\sqrt \xi(\sigma_\ell^\mathrm{TT}\sigma_\ell^\mathrm{EE})^{N/2}},
\end{equation}
where $\boldsymbol\theta=\{C_\ell^\mathrm{TT},C_\ell^\mathrm{TE},C_\ell^\mathrm{EE}\}$, $\hat c=\hat C_\ell^\mathrm{TE}$, $\rho=C_\ell^\mathrm{TE}/(\sigma_\ell^\mathrm{EE}\sigma_\ell^\mathrm{TT})$ is the correlation coefficient between the two noisy vectors,   ${\sigma_\ell^\mathrm{XX}=\sqrt{C_\ell^\mathrm{XX}+N_\ell^\mathrm{XX}}}$ is the total uncertainty on the power spectrum $C_\ell^\mathrm{XX}$, $N_\ell^\mathrm{XX}$ is the noise power spectrum, $N=2\ell+1$ is the number of modes per multipole, $\xi=(1-\rho^2)\sigma_\ell^\mathrm{TT}\sigma_\ell^\mathrm{EE}$ is a useful auxiliary variable, $\Gamma$ is the gamma function, and $K_\nu$ is the modified Bessel function of the second kind of order $\nu$. 

To better understand the variance-gamma distribution, we  show how it reduces to the $\chi^2$ distribution when taking a cross spectrum of identical vectors, i.e., $\rho\to1$. This distribution $P(x)$ is proportional to $|x|^{(N-1)/2} e^{N\rho x/\xi}K_\ell\left(\frac{N|x|}{\xi}\right)\xi^{-1/2}$. The modified Bessel function of the second kind decays exponentially, and its zeroth order expansion is given by $K_\nu(x)\approx \sqrt{\frac{\pi}{2x}}e^{-x}$~\citep{abramowitz+stegun}. In the limit of large $x$, the functional form of the variance-gamma distribution goes to
\begin{align}
    P(x)&\propto |x|^{(N-1)/2}
    \exp\left(\frac{N\rho x}{\xi}\right)
    \exp\left(-\frac{N|x|}{\xi}\right)\sqrt{\frac{\pi \xi}{2N|x|}}\xi^{-1/2}
    \\
    &\propto |x|^{N/2-1}
    \exp\left(\frac{\rho x-|x|)}{1-\rho^2}\right).
\end{align}
For perfectly correlated variables, the correlation $\rho=1$ and the data are positive definite with $x\geqslant0$, giving $P(x)\propto x^{N/2-1}e^{-x/2}$, the $\chi^2$ distribution with $N$ degrees of freedom.

This parameterization of the  variance-gamma distribution has mean and variance per multipole
\begin{align}
\langle\hat C_\ell^\mathrm{TE}\rangle&=C_\ell^\mathrm{TE}\\
\mathrm{var}(\hat C_\ell^\mathrm{TE})&=\frac1{2\ell+1}\left[(C_\ell^\mathrm{TE})^2+C_\ell^\mathrm{TT}
C_\ell^\mathrm{EE}\right],
\label{eq:varTE}
\end{align}
in agreement with the mean and variance of the off-diagonal component of the Wishart distribution and the  Gaussian distribution of $a_{\ell m}^\mathrm T$ and $a_{\ell m}^\mathrm E$. We have also validated the functional form using $10^4$ realizations of $\boldsymbol a_{\ell m}$ vectors,  and find that the distribution of $\hat C_\ell^\mathrm{TE}$ agrees with \autoref{eq:vargam}.

\subsection{Likelihood for instantaneous reionization}
\label{subsec:like_zreio}

To demonstrate the relative constraining power of the Wishart, $\chi^2$, and variance-gamma likelihoods,  we start with the theoretical power spectra as a function of the reionization optical depth $\tau$ in the case of instantaneous reionization, $C_\ell^\mathrm{TT/TE/EE}=f(\tau, A_s)$, with $A_s e^{-2\tau}$ fixed.  Additionally, we include a white noise component that is uncorrelated between $I$, $Q$, and $U$ and whose amplitude $w_p^{-1/2}$ varies between 0--230~$\mathrm{\mu K\,arcmin}$.
Using this formalism allows us to make predictions for the best-case constraining power on $\tau$ for future experiments, assuming instantaneous reionization.

\begin{figure}
    \centering
    \includegraphics[width=\columnwidth]{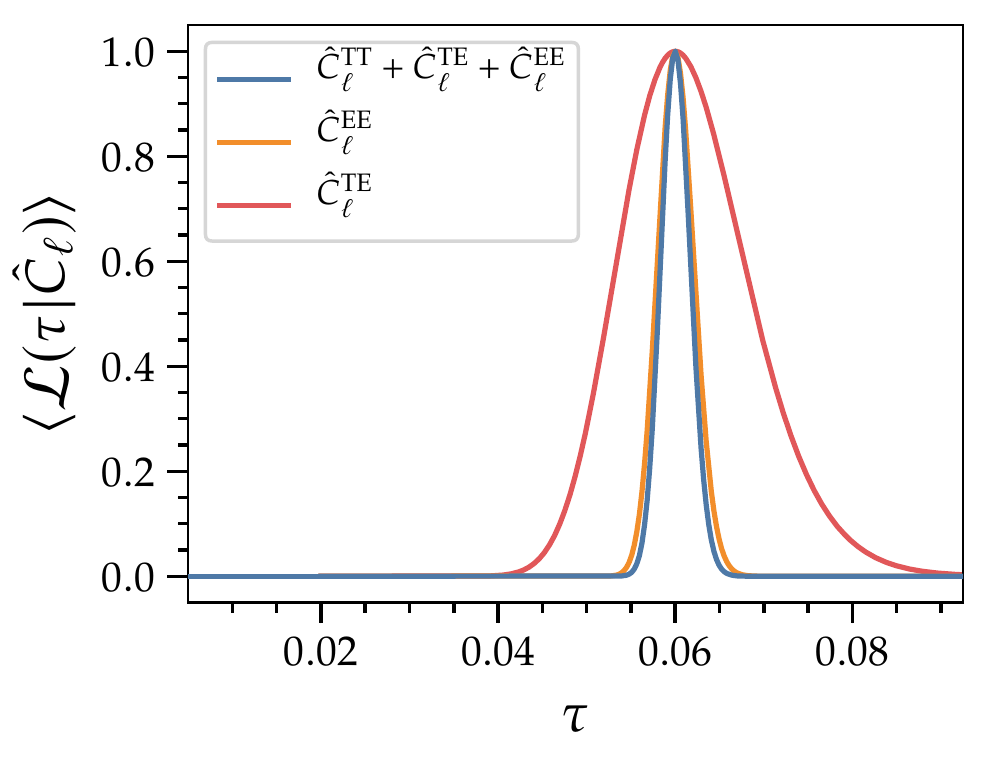}
    \caption{Normalized product of 50\,000 likelihood distributions of $\hat{\mathbf C}_\ell$ realizations with input $\tau=0.06$. 
    We plot the likelihood from the variance-gamma distribution for \TE\ (red), the likelihood from the $\chi^2$ distribution for \EE\ (orange), and the likelihood from the Wishart distribution for \TTTEEE\ (blue). The standard deviations of these distributions for input $\tau=0.06$ are  $\sigma_\tau=\{0.0072, 0.0021, 0.0017\}$ respectively. 
 }
    \label{fig:neglect_te}
\end{figure}

We characterize the likelihood of $\tau$ by evaluating $\mathcal L(\tau|\{\hat C_\ell^\mathrm{TT/TE/EE}\})$ for many realizations of the CMB sky.
We create 50\,000 realizations of $\boldsymbol a_{\ell m}$ with ${2\leqslant\ell\leqslant100}$ to test this formalism using the \texttt{HEALPix}\footnote{\textbf Hierarchical \textbf Equal \textbf Area iso\textbf Latitude \textbf{Pix}elation\newline \url{https://healpix.sourceforge.io/}} routine \texttt{synalm}.
In \autoref{fig:neglect_te}, we show the averaged likelihood of these different spectra in the case of a full-sky cosmic variance-limited measurement, and obtain $\sigma_\tau^\mathrm{TT+TE+EE}=0.0017$, $\sigma_\tau^\mathrm{EE}=0.0021$, and $\sigma_\tau^\mathrm{TE}=0.0072$. The TE-only constraint is comparable to the uncertainty from \planck, $\sigma_\tau^\mathit{Planck}=0.007$, which only includes E-mode data.
The distribution for \TE\ in \autoref{fig:neglect_te} is visibly skewed. This is a manifestation of the underlying skewed distributions that the $\hat C_\ell^\mathrm{TT/TE/EE}$ are themselves drawn from. 

\begin{figure}
    \centering
    \includegraphics[width=\columnwidth]{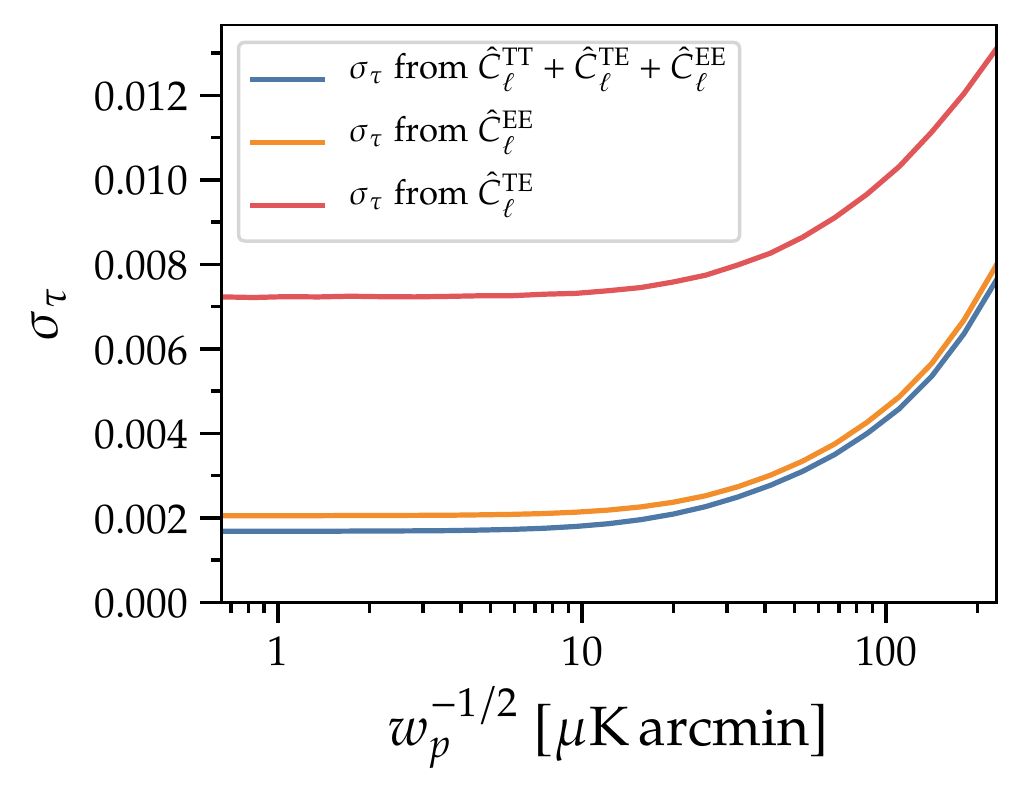}
    \caption{Uncertainty on $\tau$ as as a function of white noise amplitude in polarization for a full-sky measurement. Using $\hat C_\ell^\mathrm{TE}$ alone is always less constraining than $\hat C_\ell^\mathrm{EE}$ by a factor of $\sim3.5$. Including \TE and $\hat C_\ell^\mathrm{TT}$ data  improves the precision of a $\tau$ measurement by 20\% over using $\hat C_\ell^\mathrm{EE}$ alone.
    }
    \label{fig:compare_sigma}
\end{figure}

Since the uncertainty on $\hat C_\ell^\mathrm{TE}$ in \autoref{eq:varTE} is a function of $(C_\ell^\mathrm{EE,th}+N_\ell^\mathrm{EE})(C_\ell^\mathrm{TT,th}+N_\ell^\mathrm{TT})$, it is reasonable to ask whether there is a combination of uncertainties that makes $\sigma_\tau^\mathrm{TE}$ competitive with $\sigma_\tau^\mathrm{EE}$. \autoref{fig:compare_sigma} demonstrates that at a given polarized white noise level $w_p^{-1/2}$, 
the constraining power on $\tau$ from the $\hat C_\ell^\mathrm{TE}$ alone is a factor of $\sim3.5$ weaker than the $\hat C_\ell^\mathrm{EE}$ constraint. This means that using \TTTEEE results in  an approximately $20\%$ increase in precision compared to using \EE data alone.

The white noise temperature component is functionally negligible for this analysis. We can see this  by looking at the components of \autoref{eq:varTE} contributing to the white noise in $\hat C_\ell^\mathrm{TE}$, ${(C_\ell^\mathrm{TT,th}+N_\ell^\mathrm{TT})(C_\ell^\mathrm{EE,th}+N_\ell^\mathrm{EE})}$. The theory-noise cross-terms are comparable when $N_\ell^\mathrm{TT}C_\ell^\mathrm{EE,th}\approx N_\ell^\mathrm{EE}C_\ell^\mathrm{TT,th}$. Since $C_\ell^\mathrm{TT,th}/C_\ell^\mathrm{EE,th}\simeq10^4$ for $\ell\lesssim100$, the polarization sensitivity $w_p^{-1/2}$ would have to be $\mathcal O(10^{-2})$ times that of temperature for the temperature spectrum's white noise component to noticeably contribute to the \TE\ variation.

\section{The CMB's sensitivity to varying reionization histories}
\label{sec:fisher}

We begin discussing a specific simple model for early reionization in \autoref{subsec:reio}, then discuss quantitative forecasts in \autoref{subsec:fisher_forecasts}.
\newpage
\subsection{A simple model for early reionization}
\label{subsec:reio}

\begin{figure}
    \centering
    \includegraphics[width=\columnwidth]{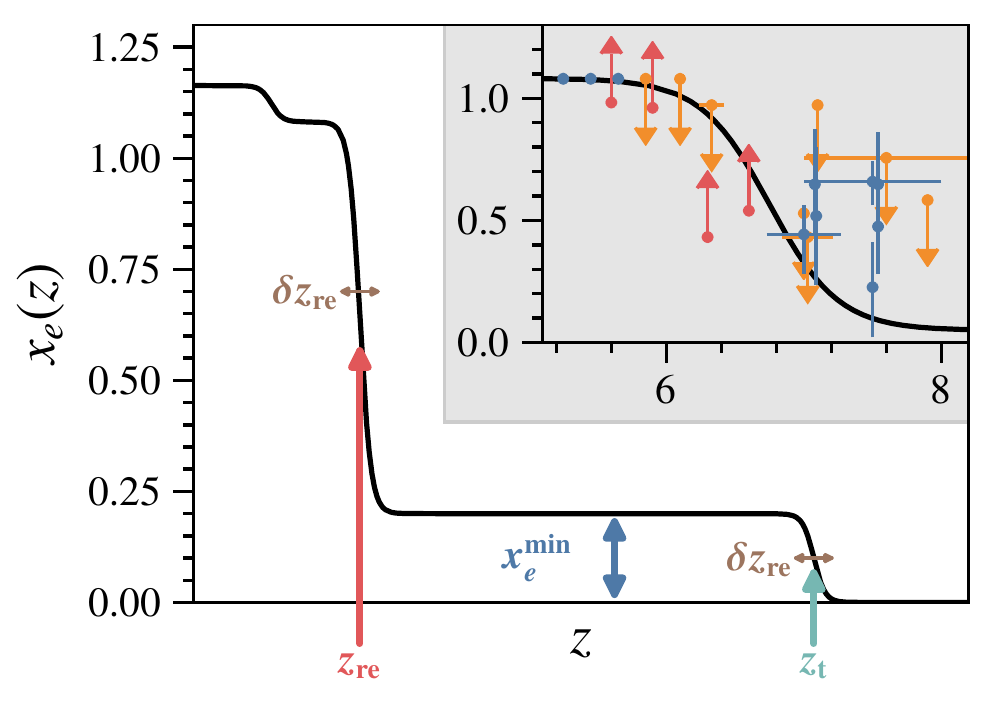}
    \caption{Visualization of our toy model for early reionization. We indicate the central redshift of late reionization $z_\mathrm{re}$ (red), the amplitude of the early reionization fraction $x_e^\mathrm{min}$ (blue), the redshift $z_\mathrm t$ where early reionization begins (cyan), and the width $\delta z_\mathrm{re}$ of these transitions (brown). We inflate $x_e^\mathrm{min}\to0.2$ to illustrate this parameter's effect in the model.
    \added{Inset: comparison of toy model with direct measurements of IGM ionization fraction. We plot measurements with upper and lower limits in blue, upper limits in orange, and lower limits in red. These data do not rule out our toy model evaluated at our fiducial parameters of $z_\mathrm{re}=6.75$ and ${x_e^\mathrm{min}=0.05}$. The data include the compilation by \citet{bouwens} and the measurements from \citet{McGreer2015}, \citet{Greig2017}, \citet{banados}, \citet{Mason2018}, \citet{davies}, \citet{Greig2019}, and \citet{Mason2019}. }
    \explain{Added inset panel to \autoref{fig:model} to compare our model with currently available data.}
    }
    \label{fig:model}
\end{figure}

We explore the constraining power of low-multipole CMB polarization data using a specific parameterization of the reionization history.
We parameterize the  global reionization history $x_e(z)$ using
the the ratio of free electrons
to hydrogen nuclei as a function of time, $x_e\equiv n_e/n_\mathrm H$,\footnote{The free electron fraction $x_e$ is greater than one at low redshifts  because of the free electrons corresponding to helium. When helium is singly ionized, the electron number density is ${n_e=n_\mathrm H+n_\mathrm{He}=(1+f_\mathrm{He})n_\mathrm H}$, and when it is doubly ionized ${n_e=(1+2f_\mathrm{He})n_\mathrm H}$.} and write the contribution to the reionization optical depth between two redshifts $z_1$ and $z_2$ as
\begin{equation}
\label{eq:tauz}
\tau(z_1,z_2)\equiv\int_{t(z_1)}^{t(z_2)}c\sigma_\mathrm T x_e\big[z(t)\big] n_\mathrm H\big[z(t)\big]\ud t.
\end{equation}
We parameterize the reionization history using  a similar model to that used in Equation A3 of \citet{heinrich},
\begin{align}
\label{eq:xe}
x_e(z)=&&\frac{1+f_\mathrm{He}-x_e^\mathrm{min}}2
&\left\{1+\tanh\left[\frac{y_\mathrm{re}-y}{\delta y}\right]
\right\}
\nonumber
\\
&&+\frac{x_e^\mathrm{min}-x_e^\mathrm{rec}}2
&\left\{1+\tanh\left[\frac{y_\mathrm t-y}{\delta y}\right]
\right\}
+x_e^\mathrm{rec}
\end{align}
where $y(z)\equiv (1+z)^{3/2}$ and $\delta y = \frac32(1+z)^{1/2}\delta z_\mathrm{re}$. The ionization fraction from recombination alone is $x_e^\mathrm{rec}$, the second transition step is given at the redshift $z_\mathrm t$, the amplitude of reionization from the second transition  is $x_e^\mathrm{min}$, and the fraction of electrons from singly ionized helium is given by $f_\mathrm{He}\equiv n_\mathrm{He}/n_\mathrm H$. 
We use this form because it parameterizes a small but nonzero early ionization fraction. An upper limit on  
$x_e(15\leq z\leq30)$
was first inferred by \citet{millea} and further constrained by \citet{planckparams18}. Figure 45 of \citet{planckparams18} shows that above $z\gtrsim10$, \planck measurements do not rule out $x_e^\mathrm{min}\approx10\%$. Motivated by this result, we choose a fiducial value of $x_e^\mathrm{min}=0.05$ to demonstrate the potential effects this ionization fraction can have on CMB measurements. We also choose $z_\mathrm{re}=6.75$ so that the total optical depth $\tau=0.06$ is consistent with the \cite{planckparams18} values.
We highlight the parameters of this model in \autoref{fig:model}, with $x_e^\mathrm{min}$ set to $0.2$ for visibility purposes. 

\added{There are several measurements of the neutral hydrogen fraction, which we summarize in the inset of \autoref{fig:model}. These direct measurements are in general agreement with the reionization constraints from CMB measurements, but do not yet provide strong evidence for or against reionization models more complex than the instantaneous tanh-like model. More concretely, these direct measurements do not rule out a free electron fraction $x_e^\mathrm{min}=0.05$ at redshifts $7\lesssim z\lesssim 8$.}\footnote{\added{NB: The highest-redshift points in \citet{Fan2006} were incorrectly transcribed as measurements and not upper limits in \citet{bouwens}. This has been corrected in \autoref{fig:model}.}}

\begin{figure*}
    \centering
    \includegraphics[width=0.8\paperwidth]{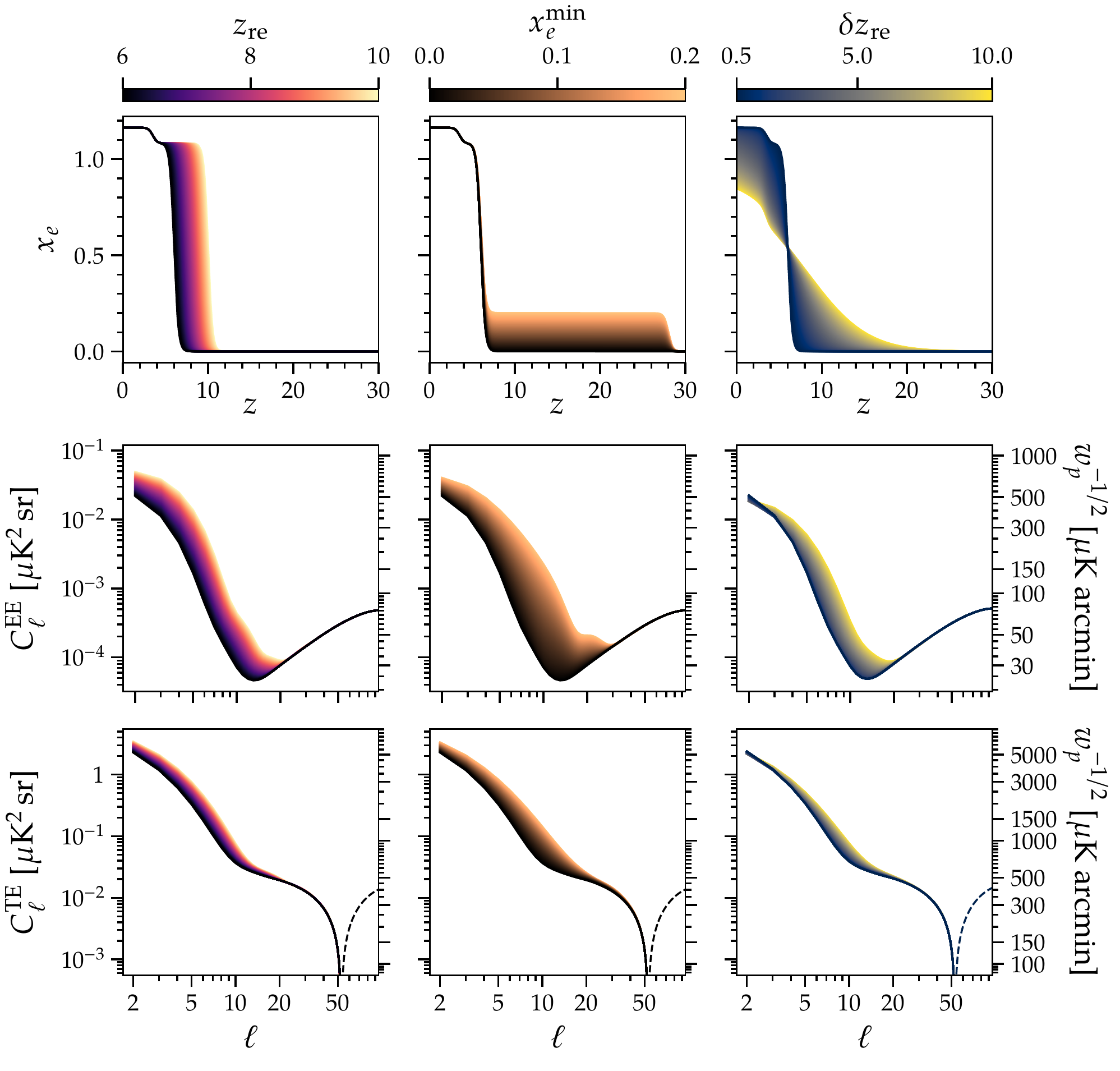}
    \caption{Reionization histories and their corresponding power spectra. Negative values are plotted as dashed lines. 
    Each column has curves with each color corresponding to a reionization history $x_e(z)$, and its corresponding spectra $C_\ell^\mathrm{EE}$ and $C_\ell^\mathrm{TE}$. In addition, the left- and right-hand axes share limits and tick values across each row.
    Each column varies $z_\mathrm{re}$, $x_e^\mathrm{min}$, and $\delta z_\mathrm{re}$ independently, with a baseline black curve $(z_\mathrm{re},x_e^\mathrm{min},\delta z_\mathrm{re})=(6,0,0.5)$, corresponding to $\tau=0.039$.
    Variations of the reionization history of the universe create corresponding variations in the polarization of the CMB at large angular scales. 
    Changing the reionization history changes both the $C_\ell^\mathrm{EE}$ and $C_\ell^\mathrm{TE}$ power spectra, and the information from these two power spectra can be used both as a better constraint on the reionization history and as a consistency check. While varying these different parameters has similar effects on the low-$\ell$ amplitude, the variation as a function of multipole can be used to eliminate degeneracies between the similar parameters.
    We include for comparison the width of reionization parameter $\delta z_\mathrm{re}$ to point out that while it is possible for this parameter to affect the CMB on large angular scales, the impact of $\delta z_\mathrm{re}$ is reduced given that the universe is ionized today, i.e., $x_e(z=6)>1$. 
    }
    \label{fig:histories}
\end{figure*}

We show the dependence of  $C_\ell^\mathrm{EE}$ and $C_\ell^\mathrm{TE}$ on these reionization histories in \autoref{fig:histories}. We choose the ranges of the parameters such that they induce roughly equivalent changes in the amplitude of the output power spectrum. 
The equivalent white noise powers are labeled on the right-hand side of \autoref{fig:histories}. 
We also vary $\delta z_\mathrm{re}$ to show that although this parameter does affect the power spectra, unphysically large widths $\delta z_\mathrm{re}\gtrsim5$ are needed to affect the power spectra as much as $z_\mathrm{re}$ and $x_e^\mathrm{min}$.
\added{Variations in $\delta z_\mathrm{re}$ induce a high-redshift tail, similar to the empirical extended phenomenological models described in \citet{Lapi2017}, \citet{Roy2018}, and \citet{Kulkarni2019}. The rightmost column of \autoref{fig:histories} shows these models are not strongly constrained by large-scale CMB polarization data.}
We fix the width of these transitions to $\delta z_\mathrm{re}=0.5$ because it is weakly constrained by E-mode power spectra for a reionization history that is complete by $z=0$.

\subsection{Constraints on high-redshift reionization}
\label{subsec:fisher_forecasts}

In the parameterization of \autoref{eq:xe}, it is natural to compare to \autoref{eq:tauz} and constrain the parameters ${\tau_{\mathrm{lo}}\equiv\tau(0, z_\mathrm{split})}$ and $\tau_{\mathrm{hi}}\equiv \tau(z_\mathrm{split},z_\mathrm{dark})$. 
We choose $z_\mathrm{dark}=100$ as a redshift sufficiently far removed from both recombination and reionization effects.
We define ${z_\mathrm{split}\equiv z_\mathrm{re}+1}$.
This parameterization essentially allows a one-dimensional mapping such that ${\tau_\mathrm{lo}=f(z_\mathrm{re})}$ and ${\tau_\mathrm{hi}=g(x_e^\mathrm{min})}$.
In the case of standard tanh-like reionization, $\tau_{\mathrm{lo}}\to\tau$ and $\tau_{\mathrm{hi}}\to0$, or equivalently $x_e^\mathrm{min}\to x_e^\mathrm{rec}$.

The primary effect of adding a second component to $x_e(z)$ is an increase in the total reionization optical depth $\tau$, and therefore the rough amplitudes of the polarized power spectra, specifically $C_\ell^\mathrm{EE}\propto\tau^2$ and $C_\ell^\mathrm{TE}\propto\tau$ at the lowest multipoles $\ell\lesssim10$.
The second and more distinguishing effect is that both of these power spectra change shape due to the different angular sizes of local quadrupoles at the primary and secondary reionization redshifts. This provides an opportunity to go beyond $\tau$ in probing the nature of reionization. We demonstrate the effects of varying $x_e^\mathrm{min}$ and $z_\mathrm{re}$ on the polarized power spectra (see \autoref{fig:histories}) using the Boltzmann code \texttt{CLASS}~\citep{CLASS}. 
For every reionization history, we compute $\tau$ and vary $A_s$ such that $A_s e^{-2\tau}$ is held constant.

Using the \texttt{CLASS} code, we set \texttt{reio\_parameterization} equal to \texttt{reio\_many\_tanh}  with $\delta z_\mathrm{re}=0.5$, fixing ${z_\mathrm t=30}$, ${f_\mathrm{He}=1.324}$ using the fiducial helium mass fraction ${Y_p=0.25}$, and ${x_e^\mathrm{rec}=2\times10^{-4}}$. We vary $z_\mathrm{re}$ and $x_e^\mathrm{min}$ to write the cosmological power spectrum as a function of two parameters,  $C_\ell^\mathrm{TT/TE/EE}=f(z_\mathrm{re}, x_e^\mathrm{min})$.

\begin{figure}
    \centering
    \includegraphics[width=\columnwidth]{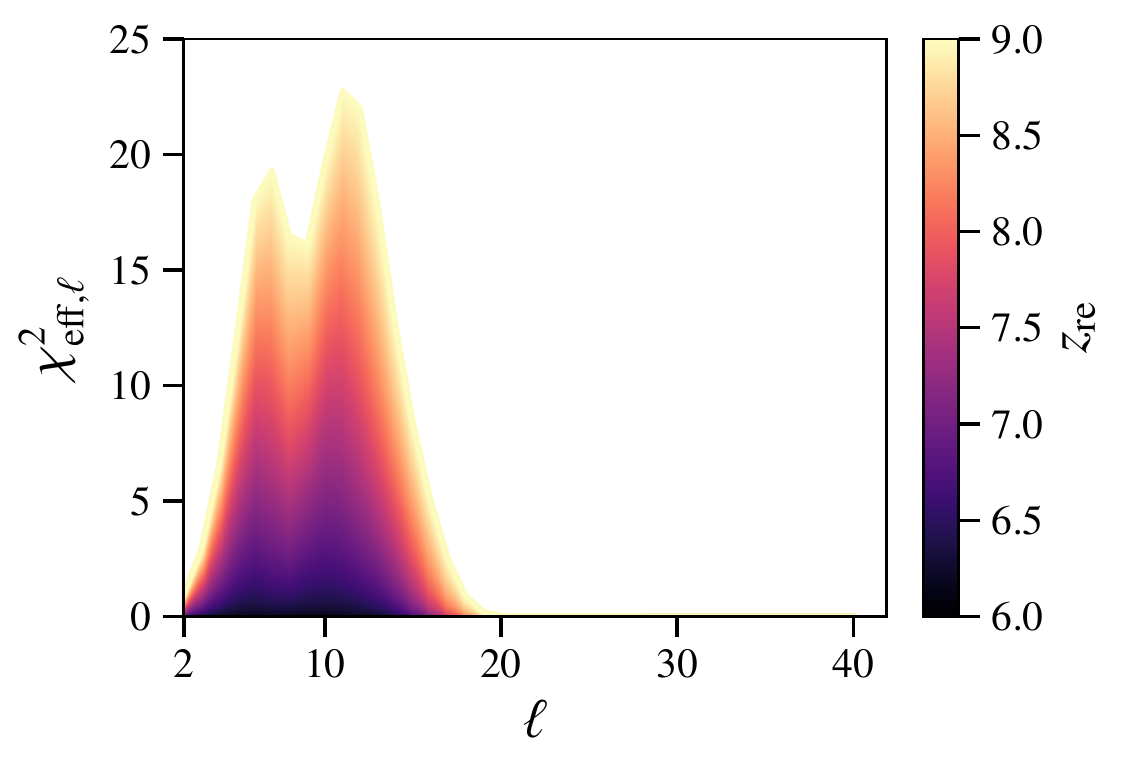}
    \includegraphics[width=\columnwidth]{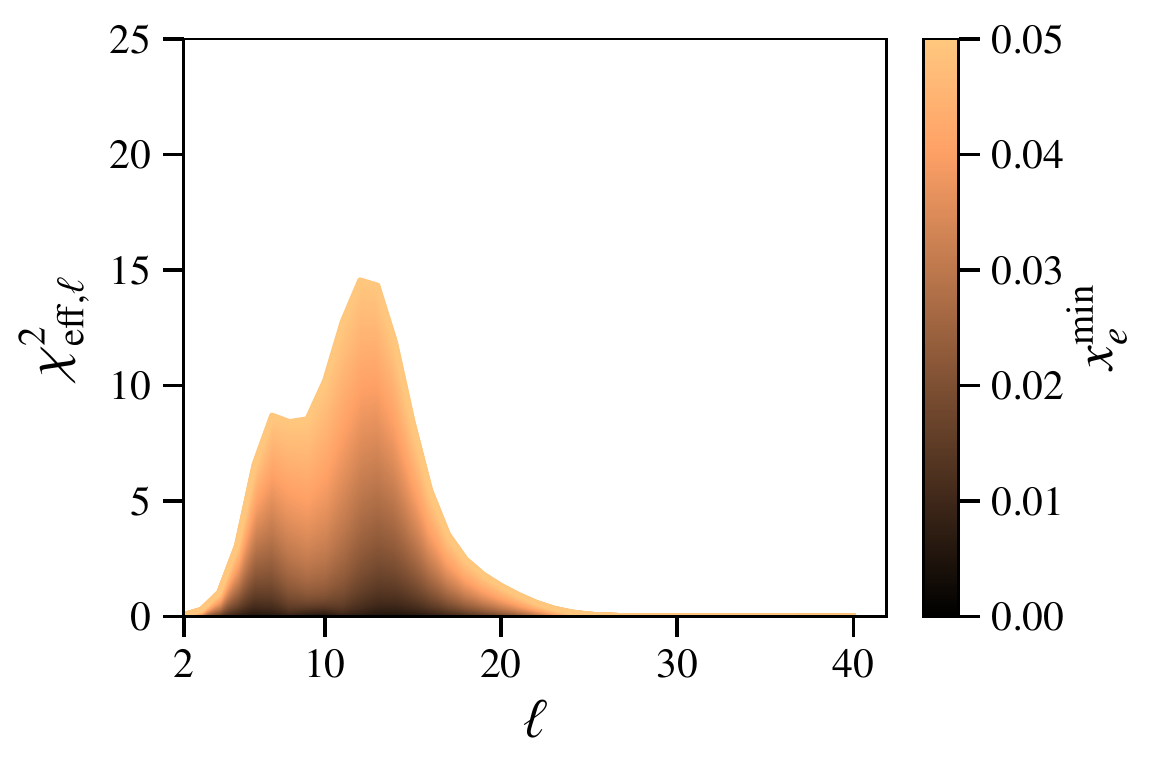}
    \caption{Effective goodness-of-fit as a function of multipole. The top subplot varies $z_\mathrm{re}$ and the bottom varies $x_e^\mathrm{min}$.
    Using an observed set of power spectra $\{\hat C_\ell^\mathrm{TT}, \hat C_\ell^\mathrm{TE}, \hat C_\ell^\mathrm{EE}\}$ that are identical to their theory values $\{C_\ell^\mathrm{TT}, C_\ell^\mathrm{TE}, C_\ell^\mathrm{EE}\}$ with $z_\mathrm{re}=6$ and $x_e^\mathrm{min}=0$, we calculate the global goodness-of-fit while varying $z_\mathrm{re}$ and $x_e^\mathrm{min}$ independently of each other.
    We note that the $2\lesssim\ell\lesssim20$ range of angular scales contains most of the effective constraining power of polarized CMB measurements. }
    \label{fig:chi2_per_ell}
\end{figure}

In \autoref{fig:chi2_per_ell}, we plot $\chi^2_\mathrm{eff,\ell}$ using \autoref{eq:chi2_eff}. By varying $z_\mathrm{re}$ and $x_e^\mathrm{min}$ separately, we can observe a few noteworthy features. First, although there is more variation in the power spectra at the very largest scales, the constraining power peaks at $\ell\simeq10$, corresponding to fluctuations on scales of tens of degrees. 
Second, the two different reionization histories have notably different $\chi^2_\mathrm{eff,\ell}$, demonstrating that the partial degeneracy between these two modifications to reionization history can be broken with high signal-to-noise measurements across this range of angular scales.
The very largest scales $\ell<10$ are much more constraining for $z_\mathrm{re}$ than $x_e^\mathrm{min}$, whereas the $10\leqslant\ell<20$ range is very sensitive to both parameters.

\begin{figure}
    \centering
    \includegraphics[width=\columnwidth]{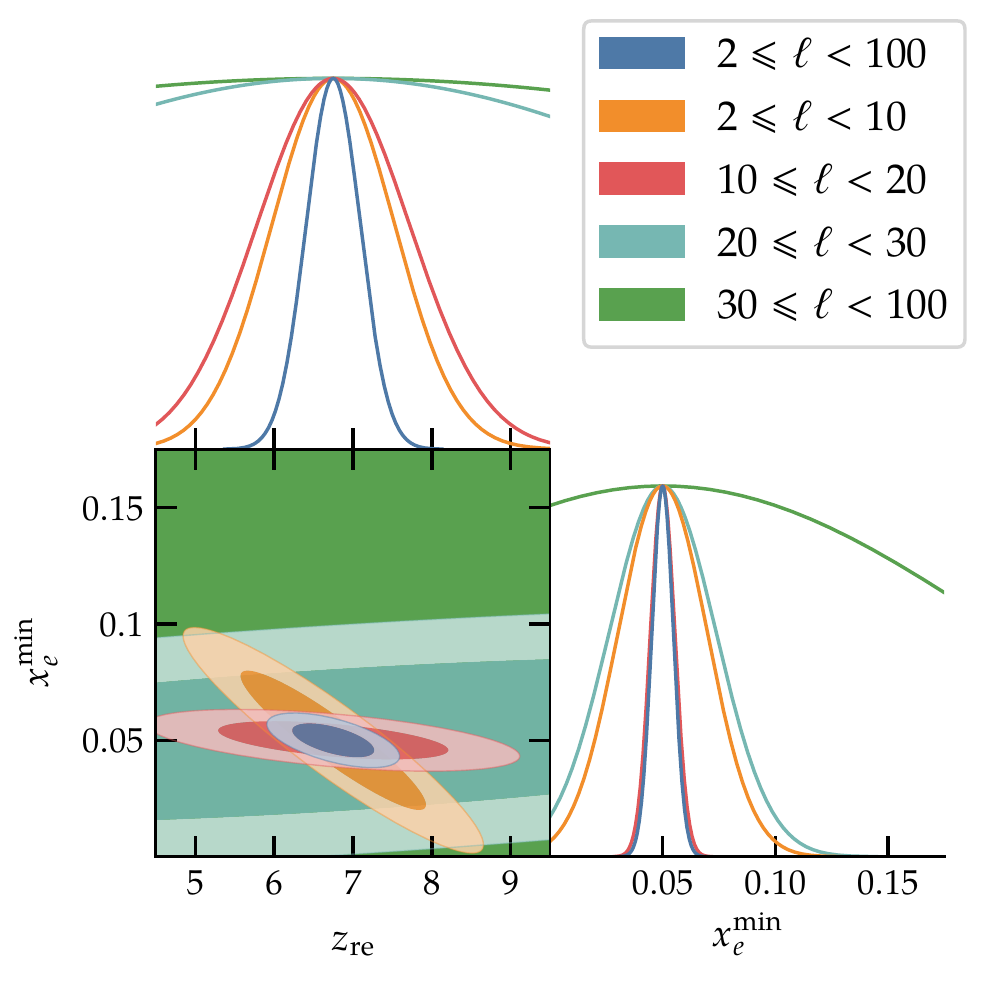}
    \caption{Fisher forecasts for cosmic variance-limited \TTTEEE data over subsets of multipole ranges. 
    The high opacity and low opacity ellipses are $1\sigma$ and $2\sigma$ contours, respectively.
    When considering the two-parameter reionization model, the $10\leqslant\ell<20$ range is most important for distinguishing between alternate models of reionization. This range has the maximum $\chi^2_\mathrm{eff}$ variation due to its relatively strong model dependence compared to higher multipoles and relatively small cosmic variance compared to lower multipoles.}
    \label{fig:contours_ellranges}
\end{figure}

\begin{figure}
    \centering
    \includegraphics[width=\columnwidth]{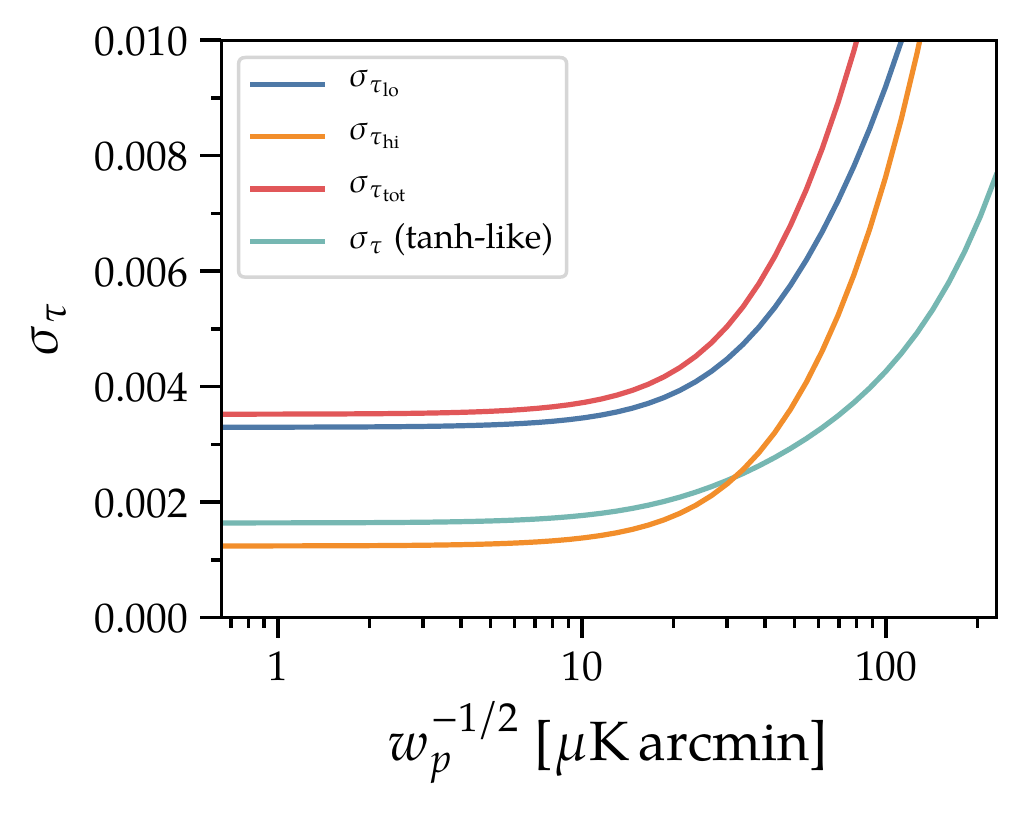}
    \caption{Constraints on $\tau$ uncertainty as a function of white noise. In all cases, the uncertainty saturates at white noise level $w_p^{-1/2}\sim 10\,\mathrm{\mu K\,arcmin}$.
    Here we display the uncertainty on the optical depth from various components of reionization, as well as the total reionization optical depth $\tau_\mathrm{tot}$. 
    }
    \label{fig:sig_versus_noise}
\end{figure}

We quantify this multipole dependence by performing Fisher forecasts on subsets of cosmic variance-limited data in  \autoref{fig:contours_ellranges}. As expected, there is relatively little constraining power in the $30\lesssim\ell\lesssim100$ multipole range, but the majority of constraining power comes from the $10\lesssim\ell\lesssim20$ range, in agreement with the shape of the $\chi^2_{\mathrm{eff},\ell}$ curves in \autoref{fig:chi2_per_ell}. While there is significant constraining power in the $2\lesssim\ell\lesssim10$ and $20\lesssim\ell\lesssim30$ ranges, the 10--20 range is by far the most important for quantitative assessment for this reionization scenario.

We show the uncertainty on the optical depth parameters $\tau_\mathrm{lo}$, $\tau_\mathrm{hi}$, $\tau_\mathrm{tot}\equiv\tau_\mathrm{lo}+\tau_\mathrm{hi}$, and $\tau$ from tanh-like reionization as a function of white noise level in \autoref{fig:sig_versus_noise}. This demonstrates that the optical depth from high-redshift reionization can be meaningfully constrained with relatively high white noise levels, and that the uncertainty on any additional optical depth from high-redshift sources can be improved by an order of magnitude  above current measurements with white noise levels as high as $10\,\mathrm{\mu K\,arcmin}$.

\begin{figure}
    \centering
    \includegraphics[width=\columnwidth]{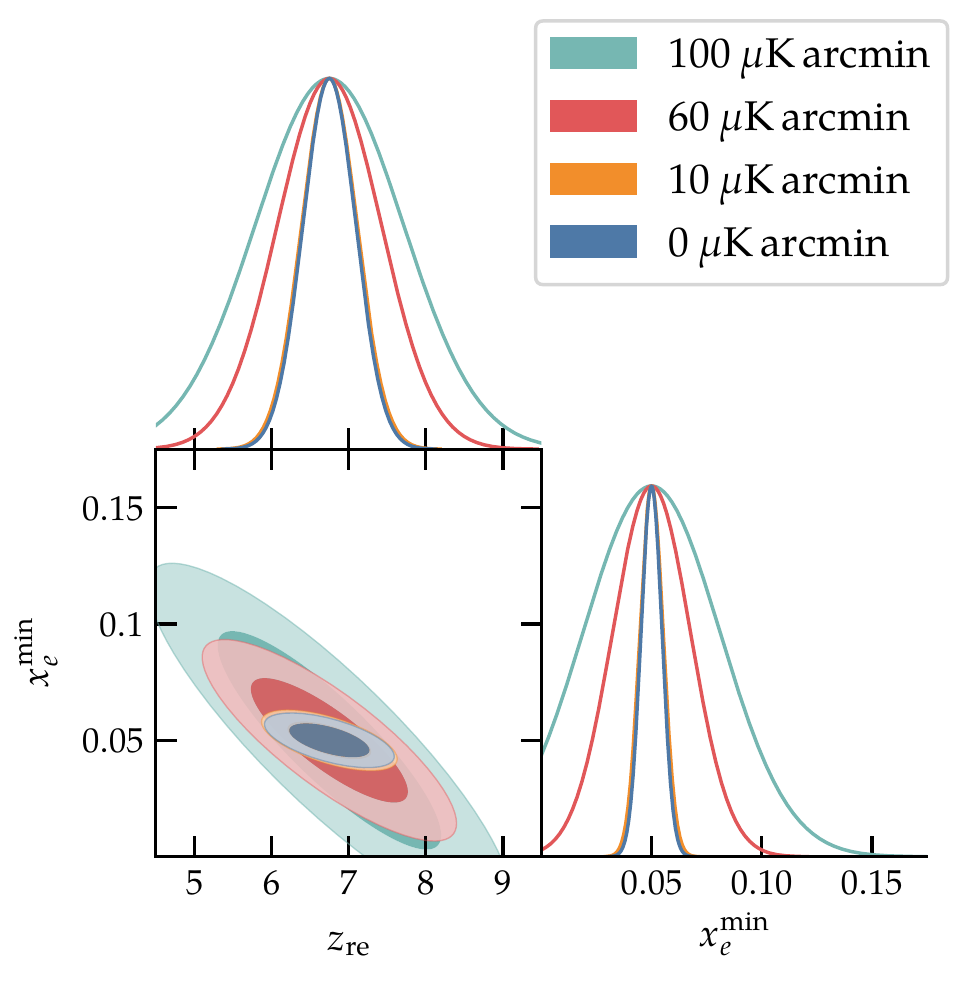}
    \caption{Fisher forecasts for $x_e^\mathrm{min}$ and $z_\mathrm{re}$ as a function of white noise level.
    The high opacity and low opacity ellipses are $1\sigma$ and $2\sigma$ contours, respectively.
    A noiseless measurement (blue) represents the fundamental limits of constraining these reionization parameters with large-scale CMB polarization measurements. For comparison, a $10\,\mathrm{\mu K\,arcmin}$ white noise level is shown (orange) and is almost completely hidden under the blue $0\,\mathrm{\mu K\, arcmin}$ contour. 
    We also plot the projected white noise contribution for a CLASS Q-band foreground cleaned map (red), and the white noise contribution in the \planck~2018 $\hat C_\ell^{100\times143}$ data (cyan).
    }
    \label{fig:contours}
\end{figure}

The constraining power of low-$\ell$ polarization data can most clearly be seen in the Fisher contours in \autoref{fig:contours}.  At current noise levels, the constraints on the reionization redshift are relatively weak, and the presence of high-redshift reionization cannot be distinguished from instantaneous reionization.  As expected, there is a negative degeneracy between $x_e^\mathrm{min}$ and $z_\mathrm{re}$ that is most pronounced at high noise levels where only the lowest multipoles contribute to the variation of the power spectra, while the degeneracy becomes less severe as the noise level decreases.

\autoref{fig:contours} demonstrates the possible advances in our understanding of reionization from the CMB. The ultimate sensitivity to $(z_\mathrm{re},x_e^\mathrm{min})=(6.75, 0.05)$---equivalent to $\tau_\mathrm{tot}=0.06$---from the CMB is shown in blue, using a Fisher forecast with zero instrumental noise. 
This noiseless measurement represents the fundamental limits for constraining these reionization parameters with large-scale CMB polarization measurements. We plot Fisher contours for noise levels ${w_p^{-1/2}=\{10, 60, 100\}\,\mathrm{\mu K\,arcmin}}$. The smallest number corresponds to the projected Cosmology Large Angular Scale Surveyor (CLASS) white noise level over 70\% of the sky in \citet{essinger-hileman} using all four of its observing bands.   This value is a benchmark for detecting primordial gravitational waves with a tensor-to-scalar ratio $r\sim0.01$, a goal for the current generation of ground-based CMB measurements.
The $w_p^{-1/2}=60\,\mathrm{\mu K\,arcmin}$ white noise level corresponds to the sensitivity of the CLASS Q-band ($40\,\mathrm{\mu K\,arcmin}$) cleaned using the \wmap K-band ($280\,\mathrm{\mu K\,arcmin}$) as a synchrotron template.
The $100\,\mathrm{\mu K\,arcmin}$ value corresponds to the geometric mean of the 100~GHz and 143~GHz white noise levels reported in Table 4 of \citet{plancklegacy}.

\begin{figure}
    \centering
    \includegraphics[width=\columnwidth]{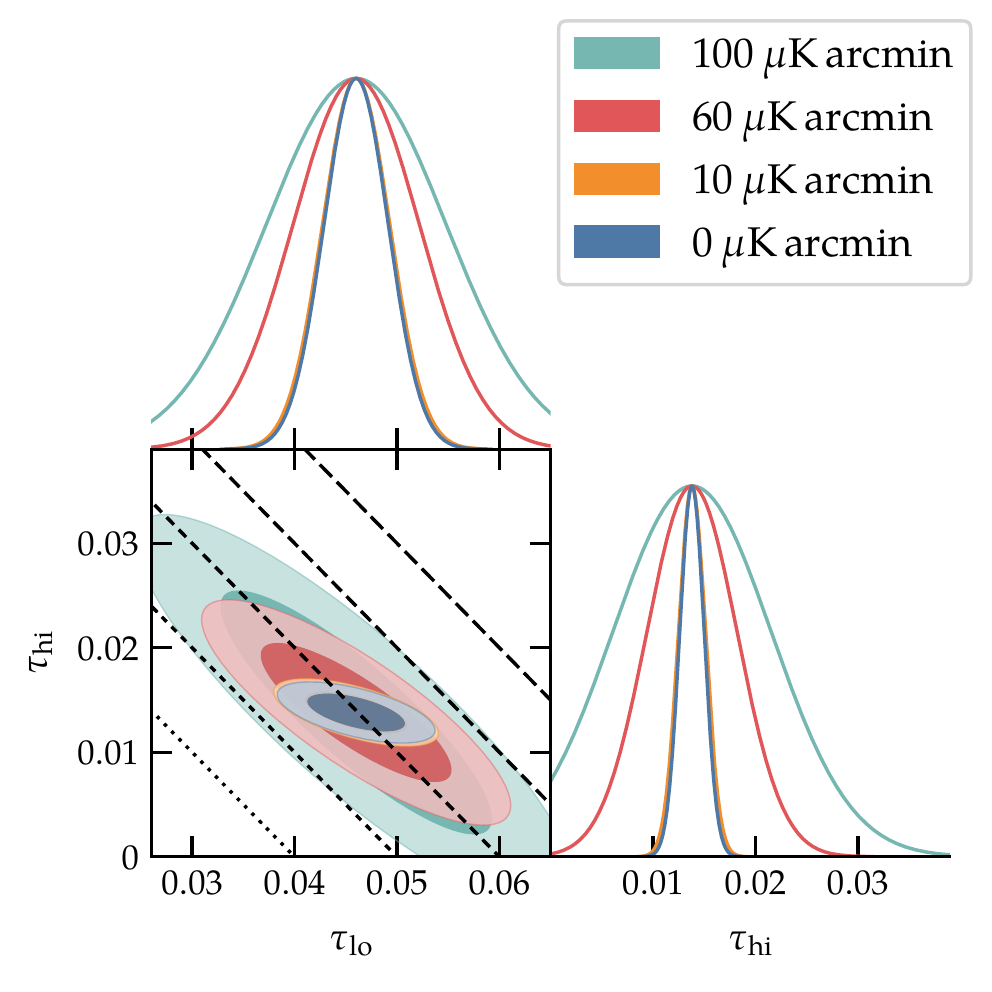}
    \caption{Fisher forecasts for $\tau_\mathrm{lo}$ and $\tau_\mathrm{hi}$ as a function of white noise level.
    The color of the contours is the same as in \autoref{fig:contours}. Additionally, we plot dashed lines of constant $\tau=\tau_\mathrm{lo}+\tau_\mathrm{hi}$ in five equal steps from $\tau=0.04$ to $\tau=0.08$, denoted by increasing dash length.
    The optimal $\tau_\mathrm{hi}$ uncertainty is $0.001$, while Fisher forecasts using $100\,\mathrm{\mu K\,arcmin}$ white noise project an uncertainty of  $\sigma_{\tau_\mathrm{hi}}=0.008$, which is degraded further by its strong negative degeneracy with $\tau_\mathrm{lo}$.
}
    \label{fig:contours_tau}
\end{figure}

We transform the contours in \autoref{fig:contours} to the integrated quantities $\tau_\mathrm{lo}$ and $\tau_\mathrm{hi}$, both approximately single-variable functions of $z_\mathrm{re}$ and $x_e^\mathrm{min}$, respectively, in
\autoref{fig:contours_tau}. We also plot lines of constant $\tau=\tau_\mathrm{lo}+\tau_\mathrm{hi}$ to show the total integrated contribution of this two-parameter reionization model.

We summarize the results of this section in \autoref{tab:vars}. We highlight data rows that are particularly constraining. This includes the full resolution cosmic variance measurement, the noiseless measurement with $10\leqslant\ell<20$, and the $w_p^{-1/2}=10\,\mathrm{\mu K\,arcmin}$ measurements. We highlight these to emphasize the relative importance of future data with these properties to constrain reionization histories.

With our fiducial parameters, the ultimate sensitivity to this model with large-scale CMB anisotropy measurements is $\sigma_{z_\mathrm{re}}=0.3$ and $\sigma_{x_e^\mathrm{min}}=0.005$. 
Remarkably, this constraint does not weaken appreciably either when examining only the multipole range $10\leqslant\ell<20$, or when the data are contaminated by white noise at the $10\,\mathrm{\mu K\,arcmin}$ level.

\begin{deluxetable}{rcrrrr}
\tablecaption{Fisher Forecasts for $z_\mathrm{re}$ and $x_e^\mathrm{min}$ when Varying Multipole Range Analyzed and White Noise Levels, Using the Fiducial Model $(z_\mathrm{re}, x_e^\mathrm{min})=(6.75,0.05)$. \label{tab:vars}}
\tablehead{$w_p^{-1/2}$  
&Multipole Range&$\sigma_{z_\mathrm{re}^{\phantom{\mathrm{min}}}}$&$\sigma_{x_e^\mathrm{min}}$&$\sigma_{\tau_\mathrm{lo}}$&$\sigma_{\tau_\mathrm{hi}}$}
\startdata
%\hline
\textbf{0} &$\boldsymbol{\phn2\leqslant\ell<100}$ & \textbf{0.3} &\textbf{ 0.005}
&\textbf{0.003}&\textbf{0.001}
\\
%\hline
0 &$\phn2\leqslant\ell<\phn10$ & 0.8&0.020&0.007&0.005
\\
%\hline
\textbf{0} &$\boldsymbol{10\leqslant\ell<\phn20}$ & \textbf{1.0}
&\textbf{0.005}&\textbf{0.009}&\textbf{0.001}
\\
%\hline
0 &$20\leqslant\ell<\phn30$ & 5.9& 0.024&0.054&0.006
\\
%\hline
0 &$30\leqslant\ell<100$ & 10.8 & 0.152&0.991&0.038
\\
%\hline
\textbf{10} & $\boldsymbol{\phn2\leqslant\ell<100}$ & 
\textbf{0.4}&\textbf{0.005}&\textbf{0.003}&\textbf{0.001}
\\
%\hline
60 & $\phn2\leqslant\ell<100$ & 0.7 &0.018&0.006&0.004
\\
%\hline
100 & $\phn2\leqslant\ell<100$ & 1.0 & 0.031&0.009&0.008
\enddata
\tablecomments{We have highlighted the three forecasts that are most constraining in bold text.}
\end{deluxetable}

\section{Conclusions}
\label{sec:conclusions}

In this work we have explored the constraining power of the CMB temperature and E-mode polarization on reionization history.
\begin{itemize}
    \item We have demonstrated the potential for a $20\%$ improvement on the precision of the reionization optical depth $\tau$ by using \TT\ and \TE\ in addition to \EE when the white noise level drops below $10\,\mathrm{\mu K\,arcmin}$.
    \item We have shown that in the case of an early 5\% ionization fraction, a scenario allowed by measurements in \citet{planckparams18}, the maximum precision from CMB large-scale measurements is $\sigma_{z_\mathrm{re}}=0.3$, and $\sigma_{x_e^\mathrm{min}}=0.005$.
    We also show that this constraint is very nearly met when the white noise level is $10\,\mathrm{\mu K arcmin}$, with $\sigma_{z_\mathrm{re}}=0.4$ and $\sigma_{x_e^\mathrm{min}}=0.005$.
    We have also shown that a key multipole range for this scenario is $10\leqslant\ell\leqslant20$, where $\sigma_{z_\mathrm{re}}=1.0$ and $\sigma_{x_e^\mathrm{min}}=0.005$.
\end{itemize}

Future measurements of the large-scale polarized CMB will be made by CLASS~\citep[currently observing]{essinger-hileman} and  \textit{LiteBIRD}\footnote{\textbf{Lite} (Light) satellite for the studies of \textbf B-mode polarization and \textbf inflation from cosmic background \textbf radiation \textbf detection}~\citep[expected launch late 2020s]{litebird}. \textit{LiteBIRD}'s goal of measuring primordial B-modes with $\sigma_r=0.001$ with ${w_p^{-1/2}=2\,\mathrm{\mu K\,arcmin}}$ observations of the whole sky will be able to constrain the reionization model presented in this paper to its cosmic variance limit. Now operating, CLASS's projected sensitivity of $w_p^{-1/2}=10\,\mathrm{\mu K\,arcmin}$ will yield observations 70\% of the sky with lower sensitivity, but as we have demonstrated here, this will be more than sufficient to constrain a period of early reionization to its cosmic variance limit.

This work has focused on the ultimate sensitivity to a specific toy model of reionization with an early high-redshift contribution. 
Processes that ionize the IGM across different epochs of cosmic time will generate different $x_e(z)$ profiles. Constraints on our model can therefore help discriminate between physical mechanisms that ionized the IGM.
Reionization history constraints from CMB measurements will both inform and complement future tomographic measurements of 21~cm emission and  of the first generation of galaxies designed to characterize $x_e(z)$.
Knowledge of the ionization history is also important for understanding the large-scale  B-modes in the reionization peak, whose fluctuations are created during the same epoch as large-scale E-modes.
Fluctuations attributed to deviations from single-field slow roll inflation can also be induced by deviations from the standard tanh-like reionization model, and these effects must be taken into account when analyzing large angular scale B-modes.

\newpage

\acknowledgments{
This research was supported in part by NASA grants NNX16AF28G, NNX17AF34G, and 80NSSC19K0526.
We acknowledge the National Science Foundation Division of Astronomical Sciences for their support of CLASS under Grant Numbers 0959349, 1429236, 1636634, and 1654494. 
Calculations for this paper were conducted using computational resources at the Maryland Advanced Research Computing Center (MARCC). 

We acknowledge the use of the Legacy Archive for Microwave Background Data Analysis (LAMBDA), part of the High Energy Astrophysics Science Archive Center (HEASARC). HEASARC/LAMBDA is a service of the Astrophysics Science Division at the NASA Goddard Space Flight Center.
We also acknowledge use of the Planck Legacy Archive. \planck\ is an ESA science mission with instruments and contributions directly funded by ESA Member States, NASA, and Canada.
Some of the results in this paper have been derived using the \texttt{healpy} and \texttt{HEALPix} packages.
This research has made use of NASA's Astrophysics Data System.

We thank Tobias A. Marriage, Zhilei Xu, Matthew Petroff, and Thomas Essinger-Hileman for discussions that improved this work.
}

\newpage

\software{
\added{\texttt{adstex}~(\url{https://github.com/yymao/adstex}),}
\texttt{camb}~\citep{cambpaper, cambcode},
\texttt{CLASS}~\citep{CLASS}, \texttt{getdist}~\replaced{\citep{getdist}}{\citep{getdist,2019arXiv191013970L}}, \texttt{healpy}~\citep{healpix, healpy}, \texttt{IPython}~\citep{ipython}, \texttt{matplotlib}~\citep{matplotlib},  \texttt{numpy}~\citep{numpy},  \texttt{scipy}~\citep{scipy} 
}

\newpage

\bibliographystyle{aasjournal}
\bibliography{reio,Planck_bib,ref_sources,test,new}

\listofchanges
\end{document}